\documentstyle[aps,prd,preprint,tighten,epsfig]{revtex}
\begin{document}
\preprint{\makebox{\begin{tabular}{r}
                        				INFNFE-06-98	\\
							BARI-TH/306-98	\\
                                                                        \\
\end{tabular}}}
\draft
\title{	Solar neutrino interactions: Using charged currents at SNO \\
	to tell neutral currents at Super-Kamiokande}
\author{	F.~L.\ Villante~${}^a$, 
		G.\ Fiorentini~${}^a$, and
		E.\ Lisi~${}^b$ \\ }
\address{$^a$Dipartimento di Fisica and Sezione INFN di Ferrara, 
             Via del Paradiso 12, I-44100 Ferrara, Italy \\
         $^b$Dipartimento di Fisica and Sezione INFN di Bari,
             Via Amendola 173, I-70126 Bari, Italy \\ }
%
\maketitle
\begin{abstract}
In the presence of flavor oscillations, muon and tau neutrinos
can contribute to the Super-Kamiokande (SK) solar neutrino signal through the
neutral current process $\nu_{\mu,\tau}\,e^-\rightarrow \nu_{\mu,\tau}\,e^-$.
We show how to separate the $\nu_e$ and $\nu_{\mu,\tau}$ event rates in SK
in a model independent way, by using the rate of the charged current
process $\nu_e\,d\rightarrow p\,p\,e^-$ from the Sudbury Neutrino Observatory
(SNO) experiment, with an appropriate choice of the SK and SNO 
energy thresholds. Under the additional hypothesis of no oscillations into
sterile states, we also show how to determine the absolute $^8$B neutrino
flux from the same data set, independently of the $\nu_e$ survival
probability. 
\end{abstract}
\medskip

\section{Introduction}

The Super-Kamiokande (SK) experiment \cite{Fu98} 
is measuring the rate $R_{\rm SK}$
of electrons produced by $^8$B solar neutrinos \cite{Ca97} through the 
scattering process
\begin{equation}
\nu_e\,e^-\rightarrow\nu_e\,e^-\ ,
\label{nue}
\end{equation}
in a fiducial mass of 22.5 kton of water and
above a threshold $T_{\rm SK}=6.5{\rm\ MeV}-m_e$ for the measured electron 
kinetic energy \cite{Su98}. There are good prospects for lowering 
$T_{\rm SK}$ to $\sim 5$ MeV in the near future \cite{Na98}.

In the presence of flavor oscillations, as suggested by various solutions
to the solar neutrino problem \cite{Ba89}, 
also muon and tau neutrinos can contribute
to $R_{\rm SK}$ through neutral current interactions
\begin{equation}
\nu_{\mu,\tau}\,e^-\rightarrow\nu_{\mu,\tau}\,e^-\ ,
\label{numutau}
\end{equation}
although it is not possible to separate the contribution of
reaction~(\ref{numutau}) from reaction~(\ref{nue}) within the SK experiment 
itself.

In this work we show how to separate in a model independent
way the $\nu_e$ and $\nu_{\mu,\tau}$ event rates in the SK experiment,
by means of the Sudbury Neutrino Observatory (SNO) \cite{Mc98} measurement of
the total electron rate ($R_{\rm SNO}$) 
from the charged current  process 
\begin{equation}
\nu_e\,d\rightarrow p\,p\,e^-\ ,
\label{nud}
\end{equation}
provided that the corresponding electron
energy threshold $T_{\rm SNO}$ is chosen appropriately.  
No other experimental information or 
theoretical assumption is required. 

More precisely,
we show that the response functions of the SK and SNO detectors happen
to be approximately equal for suitably chosen values of $T_{\rm SK}$ and
$T_{\rm SNO}$, within errors much smaller than the uncertainties associated
to the cross section for reaction~(\ref{nud}). This lucky circumstance
makes our approach truly model independent: no prior assumption is required
on the absolute $^8$B neutrino flux $\Phi_{\rm B}$, on the energy (in)dependence
of the oscillation probability, or on the presence of possible sterile
neutrinos. Furthermore, if only active neutrinos are considered, the
absolute value of $\Phi_{\rm B}$
can also be determined from the same data ($R_{\rm SK}$ and $R_{\rm SNO}$).

The plan of the paper is as follows. In Section~II we set the notation.
In Section~III we study the response functions of SK and SNO and show that
they can be equalized to a good approximation 
by tuning the energy thresholds. In Section~IV we
work out the consequences of this empirical
equality. We draw our conclusions in 
Sec.~V.

We conclude this Section by estimating the rate of 
solar-neutrino induced electrons $R_{\rm SK}$ actually occurring
in the SK detector. The rate of {\em observed\/} events above 
$T_{\rm SK}\simeq 6$ MeV is $0.60\pm 0.01({\rm stat})\pm0.02(\rm {syst})$ 
events/kton/day \cite{Su98}. However, this number is expected to be
smaller than the actual rate of $\nu e$ interactions, due to inefficiencies
in the data reduction chain. The total signal efficiency
$\varepsilon$ appears to be dominated by three cuts: (1) Noise cut, 
$\varepsilon_n = 94.2\%$ \cite{Fu98}; (2) Spallation cut, 
$\varepsilon_s = 80\%$ \cite{Su98}; and gamma ray cut, 
$\varepsilon_\gamma = 92.2\%$ \cite{Su98}.
Then we get $\varepsilon = \varepsilon_n\, 
\varepsilon_s\,\varepsilon_\gamma = 70\%$, in agreement with the signal
efficiency
quoted in \cite{Ya98}. Therefore, we estimate the total number
of $\nu e$ interactions (detected and undetected) 
occurring in SK
as:
\begin{eqnarray}
R_{\rm SK}(T_{\rm SK} = 6{\rm\  MeV})
& = & 0.60\pm 0.01 \pm 0.02\cdot \varepsilon^{-1}\\
& \simeq & 0.86 \pm  0.03 {\rm\ \ events/kton/day}\ .
\label{R1}
\end{eqnarray}
Analogously, and by cutting the electron energy spectrum \cite{Su98} 
at $T_{\rm SK}=7$ MeV, we obtain
\begin{equation}
R_{\rm SK}(T_{\rm SK} = 7{\rm\  MeV})
 \simeq  0.61\pm 0.025  {\rm\ \ events/kton/day}\ .
\label{R2}
\end{equation}
Our estimates (\ref{R1},\ref{R2}) are tentative and will be used
only for some numerical examples,
the main idea of this work being independent on the specific value
of $R_{\rm SK}$. It is understood that, when the SNO data will become
available and will be compared with the SK data, one should use 
both rates $R_{\rm SK}$ and $R_{\rm SNO}$ as corrected for 
efficiency effects by the experimental collaborations themselves.

\section{Notation}

We denote the
absolute flux of $^8$B solar neutrinos  by 
$\Phi_{\rm B}$ and its energy spectrum \cite{Ba96} by $\varphi(E_\nu)$,
\begin{equation}
\Phi_{\rm B}=\int dE_\nu\,\varphi(E_\nu)\ ,
\label{phi}
\end{equation}
$E_\nu$ being the neutrino energy. We do not assume any prior
estimate of $\Phi_{\rm B}$ from standard solar models 
\cite{Ca97,BP98}
in this work.

In the presence of neutrino oscillations, we denote the $\nu_e$ oscillation
probabilities into other states ($\nu_\mu$, $\nu_\tau$, or sterile $\nu_s$)
as
\begin{equation}
P_{e\alpha}(E_\nu)=P(\nu_e\to\nu_\alpha)  \ \ \ \ (\alpha=e,\mu,\tau,s)\ ,
\label{P}
\end{equation}
subject to the unitarity constraint $\sum_\alpha P_{e\alpha}=1$. No assumption
is made on the functional form of $P_{e\alpha}(E_\nu)$.

The neutrino cross sections for the reactions 
(\ref{nue}),
(\ref{numutau}), and
(\ref{nud}) are indicated as 
$\sigma_e$, $\sigma_{\mu\tau}$, and $\sigma_{\rm CC}$, 
respectively. It is understood that each cross section $\sigma_X$
($X=e,\,\mu\tau,\,{\rm CC}$) is corrected for the energy threshold
and resolution effects appropriate to the the SK and SNO detectors,
\begin{equation}
\sigma_X(E_\nu,T_{\rm min})=\int_{T_{\rm min}}dT\int dT'\,r(T,T')\,
\frac{d\sigma_X(E_\nu,T')}{dT'}\ ,
\label{sigmaX}
\end{equation}
where $T'$ and $T$ are the {\em true\/} and {\em measured\/} electron kinetic
energies, respectively, $T_{\rm min}$ is the energy threshold (equal to
$T_{\rm SK}$ for  $\sigma_e$ and $\sigma_{\mu\tau}$ and to $T_{\rm SNO}$ for
$\sigma_{\rm CC}$),%
\footnote{$T_{\rm SK}$ and $T_{\rm SNO}$ are {\em data analysis\/}
thresholds that can be chosen freely, provided that they are
greater than the {\em detector trigger\/} thresholds.}
$d\sigma_X/dT'$ is the differential cross section, and
$r(T,T')$ is the energy resolution function
\begin{equation}
r(T,T')=\frac{1}{\sqrt{2\pi}\Delta_{T'}}\exp\left(
-\frac{(T-T')^2}{2\Delta_{T'}^2} 
\right)\ ,
\label{r}
\end{equation}
with a one-sigma width $\Delta_{T'}$ scaling as
\begin{equation}
\Delta_{T'}=\Delta_{10}\sqrt{\frac{T'}{\rm 10\ MeV}}\ ,
\label{Delta}
\end{equation}
$\Delta_{10}$ being equal to 1.5 MeV for SK \cite{Su98,Na98} and to
1.4 MeV for SNO \cite{SNOt}. The differential cross sections are taken from
\cite{Si95} 
for reactions~(\ref{nue},\ref{numutau}) and from \cite{Ku94} for 
reaction~(\ref{nud}).

In the calculation of event rates, the cross sections
$\sigma_X$ always appear in the  
combination $\varphi\,\sigma_X$. It is then useful to define two quantities 
(related to $\varphi\,\sigma_X$) that
characterize completely the detector response to a given neutrino reaction,
namely, the energy averaged cross section $\overline\sigma_X$,
\begin{equation}
\overline\sigma_X(T_{\rm min}) = 
\frac{\displaystyle\int dE_\nu\,\varphi\,\sigma_X}
{\displaystyle\int dE_\nu\,\varphi}\ ,
\label{barsigmaX}
\end{equation}
and the normalized response function $\varrho_X$,
\begin{equation}
\varrho_X(E_\nu,T_{\rm min})=\frac{\varphi\,\sigma_X}{\displaystyle
\int dE_\nu\,\varphi\,\sigma_X}
\label{rhoX}
\end{equation}
$(X=e,\mu\tau,{\rm CC})$.
We remark that both functions $\overline\sigma_X(T_{\rm min})$ and
$\varrho_X(E_\nu,T_{\rm min})$ do not depend on the value of
$\Phi_{\rm B}$ nor on $P_{e\alpha}(E_\nu)$; they are completely
determined from detector properties, cross sections and $^8$B decay
spectrum.

Given the above definitions, the rate of electrons produced
per unit time and target electron in Super-Kamiokande through reactions
(\ref{nue}) and (\ref{numutau}) can be generally written as
\begin{equation}
R_{\rm SK}^e(T_{\rm SK})=\Phi_{\rm B}\, \overline\sigma_e \int dE_\nu \, 
\varrho_e\, P_{ee}
\label{Re}
\end{equation}
and 
\begin{equation}
R_{\rm SK}^{\mu\tau}(T_{\rm SK})=\Phi_{\rm B}\, \overline\sigma_{\mu\tau} 
\int dE_\nu \, 
\varrho_{\mu\tau}\, (1-P_{ee}-P_{es}), 
\label{Rmutau}
\end{equation}
respectively, having in mind that the SK detector does not measure 
$R^e_{\rm SK}$ and $R^{\mu\tau}_{\rm SK}$ separately, but only their sum,
\begin{equation}
R_{\rm SK}=R^e_{\rm SK}+R^{\mu\tau}_{\rm SK}\ .
\label{RSK}
\end{equation}

Analogously, the rate of electrons produced per unit time and target 
deuterons in the SNO detector through the charged current reaction~(\ref{nud})
reads
\begin{equation}
R_{\rm SNO}(T_{\rm SNO})=\Phi_{\rm B}\, \overline\sigma_{\rm CC} \int dE_\nu \, 
\varrho_{\rm CC}\, P_{ee}\ .
\label{RSNO}
\end{equation}
We stress that Eqs.~(\ref{Re}--\ref{RSNO}) are written in the most general
way and with no approximation.%
\footnote{
It is understood that the experimental rates
to be compared with such equations must be corrected for the
detector efficiencies, as also remarked in the Introduction.}

A final remark about units.
If $\Phi_{\rm B}$ is given in
cm$^{-2}$s$^{-1}$ and the cross sections
in cm$^2$, then the rates $R$ are expressed in units of events per second per
target particle (electrons in SK and deuterons in SNO), 
equivalent to $10^{36}$ generalized 
``Solar Neutrino Units'' (SNU's). Given that the molecular weight of water
(heavy water) is 18 g/mol (20 g/mol),
one SNU  corresponds to 28.9
event/kton/day in SK (5.2 event/kton/day in SNO).

\section{Cross sections and response functions for SK and SNO}

Figure~1 shows the energy averaged cross sections $\overline\sigma_e$
and $\overline\sigma_{\mu\tau}$ as a function of the SK threshold
$T_{\rm SK}$, as well as $\overline\sigma_{\rm CC}$ as a function of the
SNO threshold $T_{\rm SNO}$. We remind that such cross sections
include the effect of the detector energy resolution 
[see Eqs.~(\ref{sigmaX},\ref{barsigmaX})]. 

Figure~2 shows the normalized response functions of SK
($\varrho_e$ and $\varrho_{\mu\tau}$) and SNO ($\varrho_{\rm CC}$)
as a function of the neutrino energy, for representative values
of the detector thresholds. 
As shown in the previous Section, the quantities in Figs.~1 and 2
characterize completely the response of SK to $\nu e$ scattering
and of SNO to $\nu_e d$ absorption. 

From Fig.~2, one can see
that the response functions 
$\varrho_e$ and $\varrho_{\mu\tau}$ are almost coincident; for any
practical purpose, one can assume that
\begin{equation}
\varrho_e(E_\nu,T_{\rm SK})=\varrho_{\mu\tau}(E_\nu,T_{\rm SK})
\label{eqSK}
\end{equation}
to a very good approximation. This is not surprising, 
since the cross  sections for $\nu_e\,e$ and $\nu_{\mu,\tau}e$ 
scattering have a similar shape in the range probed by SK, 
up to an overall factor \cite{Hi87}. It is more intriguing
to notice that the SK and SNO response functions in Fig.~2
happen to be very similar, provided that the SNO threshold is chosen
about 2 MeV below the SK threshold. It appears that, using
different thresholds
and with the help of the energy resolution
smearing, the differences in the SK and SNO cross sections (weighted
by the $^8$B neutrino spectrum) can be largely compensated.

For each fixed value of $T_{\rm SK}$, we
have then maximized the agreement between
 $\varrho_e(E_\nu,T_{\rm SK})$ and $\varrho_{\rm CC}(E_\nu,T_{\rm SNO})$
by tuning $T_{\rm SNO}$, so as to minimize 
the integral reminder
\begin{equation}
\delta=\int dE_\nu |\varrho_e - \varrho_{\rm CC}|\ .
\label{delta}
\end{equation}
The results are shown in Fig.~3 for some representative values of the
thresholds. One sees that the 
two response functions $\varrho_e$ and $\varrho_{\rm CC}$ 
can be equalized to a good approximation. 
Indeed, we find that the reminder $\delta$ is always $\lesssim 0.04$.

We find that the values of
$(T_{\rm SK},T_{\rm SNO})$ that minimize $\delta$ satisfy
the approximate
relation $T_{\rm SNO} = 0.995\, T_{\rm SK} - 1.71$ (MeV), with
an accuracy sufficient for practical purposes. Of course, if the true
SNO energy resolution function
turns out be different from our prospective shape 
[see Eqs.~(\ref{r},\ref{Delta})], this relation can also change
slightly.
The comparison of the SK and SNO response functions should be finalized
when the SNO detector will be operated and calibrated.

In short, in the calculations of the electron rates $R_{\rm SK}$ and
$R_{\rm SNO}$ one can take
\begin{equation}
\varrho_{\mu\tau}(E_\nu,T_{\rm SK})=
\varrho_e(E_\nu,T_{\rm SK}) = \varrho_{\rm CC}(E_\nu,T_{\rm SNO})
\label{eqSNO}
\end{equation}
within an accuracy of a few percent or less, provided that
the SK and SNO thresholds obey the empirical relation
\begin{equation}
T_{\rm SNO} = 0.995\, T_{\rm SK} - 1.71 {\rm\ \ (MeV)}\ .
\label{thresh}
\end{equation}
The model-independent consequences of the two equations above
will be worked out in the next Section.

We point out that the comparison of the response functions
of solar neutrino experiments have always provided useful insights in
the interpretation of the solar neutrino problem. For
instance, an empirical equality between the response functions
of the Homestake and Kamiokande experiments to the $^8$B neutrino
flux was used in \cite{Ba91} 
and later in \cite{Kw94} to make some model-independent
statements on the $^8$B flux suppression. An equality similar to
Eq.~(\ref{eqSK}) was used in \cite{Hi87} 
to derive a lower bound on the $^8$B
flux. Even {\em inequalities\/} between response
functions have been used to study or constrain  the neutrino oscillation
probabilities  \cite{Kw95}. 
However, to our knowledge, Eqs.~(\ref{eqSNO},\ref{thresh})
have not been derived prior to this work.

\section{Model independent relations}

In this section we derive the model-independent consequences
of the empirical equality (\ref{eqSNO}), which holds with a precision of
a few percent or less, when $T_{\rm SNO}$ and $\rm T_{SK}$
satisfy Eq.~(\ref{thresh}). We also discuss the associated uncertainties.

From Eqs.~(\ref{Re}--\ref{RSNO}) and
Eqs~(\ref{eqSNO},\ref{thresh})
one easily derives a first important relation,
\begin{equation}
\frac{R^{\mu\tau}_{\rm SK}(T_{\rm SK})} 
{R_{\rm SK}(T_{\rm SK})} = 1- 
\frac{R_{\rm SNO}(T_{\rm SNO})}{R_{\rm SK}(T_{\rm SK})}\,
\frac{\overline\sigma_e(T_{\rm SK})}{\overline\sigma_{\rm
CC}(T_{\rm SNO})}
\label{I}
\end{equation}
which allows to determine in a model independent
way the fractional contribution of (purely neutral
current)
$\nu_{\mu,\tau}$ interactions to the total
$\nu e$ scattering rate $R^{\mu\tau}_{\rm SK}$
in Super-Kamiokande. 
We stress that the above equation 
{\em does not depend\/}
on either $\Phi_{\rm B}$ or 
the probability functions $P_{e\alpha}=P_{e\alpha}(E_\nu)$;
in particular, it holds also for nonzero mixing with
a sterile neutrino, $P_{es}(E_\nu)\neq 0$.

Under the hypothesis of no oscillations
into sterile states, $P_{es}=0$, 
a second important relation can be derived from
Eqs.~(\ref{Re}--\ref{RSNO}) and
Eqs~(\ref{eqSNO},\ref{thresh}):
\begin{equation}
\Phi_{\rm B}= \frac{1}{\overline\sigma_{\mu\tau}(T_{\rm SK})}
\left[ R_{\rm SK}(T_{\rm SK})- R_{\rm SNO}(T_{\rm SNO})
\,\frac{\overline\sigma_e(T_{\rm SK})-\overline\sigma_{\mu\tau}(T_{\rm SK})}
{\overline\sigma_{\rm CC}(T_{\rm SNO})}
\right]\ .
\label{II}
\end{equation}
The above equation gives $\Phi_{\rm B}$  independently of the
functional form of $P_{ee}(E_\nu)$.

Equations~(\ref{I}) and (\ref{II}) represent the main results of our work.
A supplementary 
relation can be derived from Eqs.~(\ref{Re}--\ref{RSK}) and 
(\ref{eqSK}), under the hypothesis
that $P_{es}=0$ {\em and\/}
that the boron neutrino flux $\Phi_{\rm B}$
is known independently (e.g., from standard
solar models):
\begin{equation}
\frac{R^{\mu\tau}_{\rm SK}}
{R_{\rm SK}}
=1-\frac{\overline\sigma_e}{\overline\sigma_e-
\overline\sigma_{\mu\tau}}\,\frac{R_{\rm SK}-\Phi_{\rm B}
\overline\sigma_{\mu\tau}}{R_{\rm SK}}
\label{III}
\end{equation}
The above equation makes use of SK data only ($R_{\rm SK}$) and does
not depend on $P_{ee}(E_\nu)$; however, its uselfuness is limited
by the requirement of a
prior knowledge of $\Phi_{\rm B}$. Using a recent standard solar
model calculation of $\Phi_{\rm B}$ \cite{BP98} and the latest
SK data \cite{Na98}
the  term subtracted from unity
in Eq.~(\ref{III})  amounts to $\sim 0.79$.
We will use this model-dependent value only
for some prospective error estimates, as we 
discuss in the following.

In order to extract $R^{\mu\tau}_{\rm SK}/R_{\rm SK}$ by means
of Eq.~(\ref{I}), one has to consider both theoretical and experimental
errors.

 Theoretical errors are of two different types:
(i) Uncertainties in the cross section ratio
$\overline\sigma_e/\overline\sigma_{\rm CC}$; and (ii) Approximations implicit
in Eq.~(\ref{eqSNO}). The first are dominated by the overall 
normalization of the CC cross section for $\nu_e d$
absorption, whose  uncertainty is about $\pm 10\%$ \cite{Ku94}. 
Concerning the approximations implicit in Eq.~(\ref{eqSNO}) one should note
 that,
if $\varrho_{\rm CC}-\varrho_{e}\neq 0$, then the right-hand side
of Eq.~(\ref{I}) acquires an additional ``error term''
equal to  $\Delta=(\Phi_{\rm B}\overline\sigma_e/R_{\rm SK})
\int dE_\nu \,P_{ee}
(\varrho_{CC}-\varrho_{e})$. The integral reminder $\delta$ (see
 Eq.~(\ref{delta})) gives an upper limit to $\delta_P =|\int dE_\nu \,P_{ee}
(\varrho_{CC}-\varrho_{e})|$, as $P_{ee}\leq 1$. Actually, 
$\delta_P$ is much smaller than $\delta$
in several oscillation cases of fenomenological interest 
(e.g., for both vacuum and
matter enhanced solutions to the solar
neutrino problem).  We have checked that 
$\delta_P \sim 6 \cdot 10^{-3}$ in the worst cases. 
If standard solar models are not too wrong,
the factor $\Phi_{\rm B}\overline\sigma_e/R_{\rm SK}$ in $\Delta$
is in the range $\sim 2$--$3$ 
(as the rate observed in SK is about 1/2--1/3 than
rate expected). Therefore, we estimate
that the approximations implicit in Eq.~(\ref{eqSNO})
introduce at most an error $\Delta \sim 0.02$ in equation Eq.~(\ref{I}).
Being much smaller than the uncertainty (i), this error
can be neglected. We also checked that this conclusion holds 
when allowance is made for variations of the SK (SNO)
energy resolution function within the quoted \cite{Na98}
(prospected \cite{SNOt}) errors.

Concerning the experimental uncertainties
 in Eq.~(\ref{I}), they are associated 
with the term $R_{\rm SNO}/R_{\rm SK}$. The 
present fractional error of $R_{\rm SK}$ is 3--4\% 
[Eq.~(\ref{R1},\ref{R2})].
However, the fractional error for
$R_{\rm SNO}$ cannot be precisely evaluated before SNO starts operation
and its background is measured.
If we assume, conservatively, a total (SNO+SK) uncertainty of $\sim 10\%$
for $R_{\rm SNO}/R_{\rm SK}$, then the total 
fractional error of the subtracted term in Eq.~(\ref{I}) is 
$\sim 15\%$ (theoretical and experimental errors added in quadrature). 
Of course, the central value of the subtracted term 
(i.e., of the $\nu_e$ contribution to the total SK rate) can only
be guessed at present. If we take the value $\sim 0.76$ from the discussion
following Eq.~(\ref{III}), then  $R^{\mu\tau}_{\rm SK}/R_{\rm SK}\simeq
1-0.79(1\pm0.15)\simeq 0.21\pm0.12$, implying that the $\nu_{\mu,\tau}$
signal can be extracted at $\sim 2\sigma$ from Eq.~(\ref{I}).

Concerning the estimate of $\Phi_{\rm B}$ from Eq.~(\ref{II}), 
 one expects 
an uncertainty of about 20\% from the same arguments. 
This value is comparable to the uncertainty affecting the theoretical
estimates of $\Phi_{\rm B}$ from solar models \cite{Ca97,BP98} 
and to the expected error
of the neutral current event rate in SNO \cite{Mc98} (which also provides
$\Phi_{\rm B}$ in the absence of sterile neutrino oscillations \cite{Ch85}). 
Therefore,
Eq.~(\ref{II}) provides us with a competitive, independent estimate of the
the boron neutrino flux. Of course, all these error estimates
should be finalized when the actual data from
SK and SNO will be compared.

Finally, we remind that
Eqs.~(\ref{I},\ref{II}) hold when the cross sections are
expressed in cm$^2$, the rates in events per target particle per second,
and $\Phi_{\rm B}$ in cm$^{-2}$s$^{-1}$. We think it useful to 
rewrite and summarize our results by using also the following
units: $[R_{\rm SK}]=[R_{\rm SNO}]=
{\rm kton}^{-1}{\rm d}^{-1}$, $[\Phi_{\rm B}]={\rm cm}^{-2}{\rm s}^{-1}$,
and $[\sigma_{X}]={\rm cm}^2$. Then, independently of 
the functional form of  the oscillation probabilities
$P_{e\alpha}(E_\nu)$, one has:
\begin{equation}
{\rm for\ any\ }\Phi_{\rm B}{\rm\ and\ }P_{es},\ \  \ \ \ 
\frac{R^{\mu\tau}_{\rm SK}}
{R_{\rm SK}}
=1 - 5.56\,
\frac{R_{\rm SNO}}
{R_{\rm SK}}\,
\frac{\overline\sigma_e}
{\overline\sigma_{\rm CC}}\ ,
\label{sum1}
\end{equation}
and
\begin{equation}
{\rm for\ }P_{es}=0,\ \ \ \ \ 
\Phi_{\rm B}=
\frac{10^{-36}}{\overline\sigma_{\mu\tau}}
\left(\frac{R_{\rm SK}}{28.9}-
\frac{R_{\rm SNO}}{5.2}\,
\frac{\overline\sigma_{e}-
\overline\sigma_{\mu\tau}}{\overline\sigma_{\rm CC}}
\right)\ ,
\label{sum2}
\end{equation}
provided that the SK and and SNO thresholds obey the
empirical Eq.~(\ref{thresh}). For these joint values of 
thresholds, we have tabulated the relevant
cross sections in Table~I.

 We conclude with a numerical example. Choosing a threshold 
$T_{\rm SK}\simeq 7.0$ MeV for SK, the corresponding observed event rate
$R_{\rm SK}$ is estimated to be
0.61 events/kton/day [Eq.~(\ref{R2})].
The SNO threshold appropriate for
comparison with SK is $T_{\rm SNO}=5.25$ MeV [see Eq.~(\ref{thresh})].
 If SNO measures,
say, 8 events/kton/day above such threshold, then one
obtains, using the cross
sections in Table~I, $R^{\mu\tau}_{\rm SK}/R_{\rm SK}\simeq0.29$
[for any $P_{es}$, Eq.~(\ref{sum1})]
 and $\Phi_{\rm B} \simeq 6.6 \cdot 10^{6}{\rm cm}^{-2}{\rm s}^{-1}$
[for $P_{es}=0$, Eq.~(\ref{sum2})].

\section{Conclusions}

We have found that an  approximate equality holds
between the Super-Kamiokande
and SNO (charged current) response functions [Eq.~(\ref{eqSNO})], 
provided that their thresholds
$T_{\rm SK}$ and $T_{\rm SNO}$ are chosen appropriately [Eq.~(\ref{thresh})]. 
 We have taken advantage of this  property, by showing
that one needs only two data (the total electron rates 
$R_{\rm  SK}$ and $R_{\rm SNO}$) to determine, in a model independent
 way, two important quantities: (a) The $\nu_{\mu,\tau}$ contribution to
the SK signal, even in the presence of additional
mixing with a sterile neutrino
[Eq.~(\ref{sum1})]; and
(b) The absolute boron neutrino flux, in the absence of oscillations into
sterile states [Eq.~(\ref{sum2})].  Therefore, the measurement of the total 
charged current event rate in SNO appears to be very interesting and
informative
in itself, and not only in relation with the neutral current measurement
to be performed in the same experiment.

\acknowledgments

	E.L.\ thanks the Department\ of Physics at the Tokyo Metropolitan
University for kind hospitality during the preparation of this work.


\begin{table}
\caption{Values of the SK cross sections $\overline\sigma_{e}(T_{\rm SK})$ 
and $\overline\sigma_{\mu\tau}(T_{\rm SK})$
and of the SNO cross section  $\overline\sigma_{\rm CC}(T_{\rm SNO})$,
for kinetic energy thresholds satisfying the relation
$T_{\rm SNO}=0.995\, T_{\rm SK}-1.71$ MeV.}
\begin{tabular}{ccccc}
$T_{\rm SK}$ & $T_{\rm SNO}$ & $\overline\sigma_{e}$ &
$\overline\sigma_{\mu\tau}$ & $\overline\sigma_{\rm CC}$ \\
(MeV) & (MeV) & ($10^{-44}$ cm$^{2})$ & ($10^{-45}$ cm$^{2}$) &
($10^{-42}$ cm$^{2}$) \\
\tableline
6.0 & 4.26 & 1.238 & 1.918 & 0.999  \\
6.2 & 4.46 & 1.147 & 1.772 & 0.971  \\
6.4 & 4.66 & 1.060 & 1.634 & 0.942  \\
6.6 & 4.86 & 0.977 & 1.504 & 0.911  \\
6.8 & 5.06 & 0.899 & 1.381 & 0.879  \\
7.0 & 5.25 & 0.825 & 1.265 & 0.845  \\
7.2 & 5.45 & 0.755 & 1.156 & 0.811  \\
7.4 & 5.65 & 0.690 & 1.054 & 0.776  \\
7.6 & 5.85 & 0.628 & 0.959 & 0.740  \\
7.8 & 6.05 & 0.571 & 0.870 & 0.703  \\
8.0 & 6.25 & 0.517 & 0.787 & 0.666  \\
8.2 & 6.45 & 0.467 & 0.710 & 0.629  \\
8.4 & 6.65 & 0.421 & 0.639 & 0.592  \\
8.6 & 6.85 & 0.378 & 0.573 & 0.555  \\
8.8 & 7.05 & 0.339 & 0.513 & 0.519  \\
9.0 & 7.25 & 0.302 & 0.457 & 0.483  \\
9.2 & 7.44 & 0.269 & 0.406 & 0.447  \\
9.4 & 7.64 & 0.239 & 0.360 & 0.413  \\
9.6 & 7.84 & 0.211 & 0.318 & 0.380  \\
9.8 & 8.04 & 0.186 & 0.280 & 0.347  \\
10.0 & 8.24 & 0.163 & 0.245 & 0.316  
\end{tabular}
\end{table}



\begin{figure}
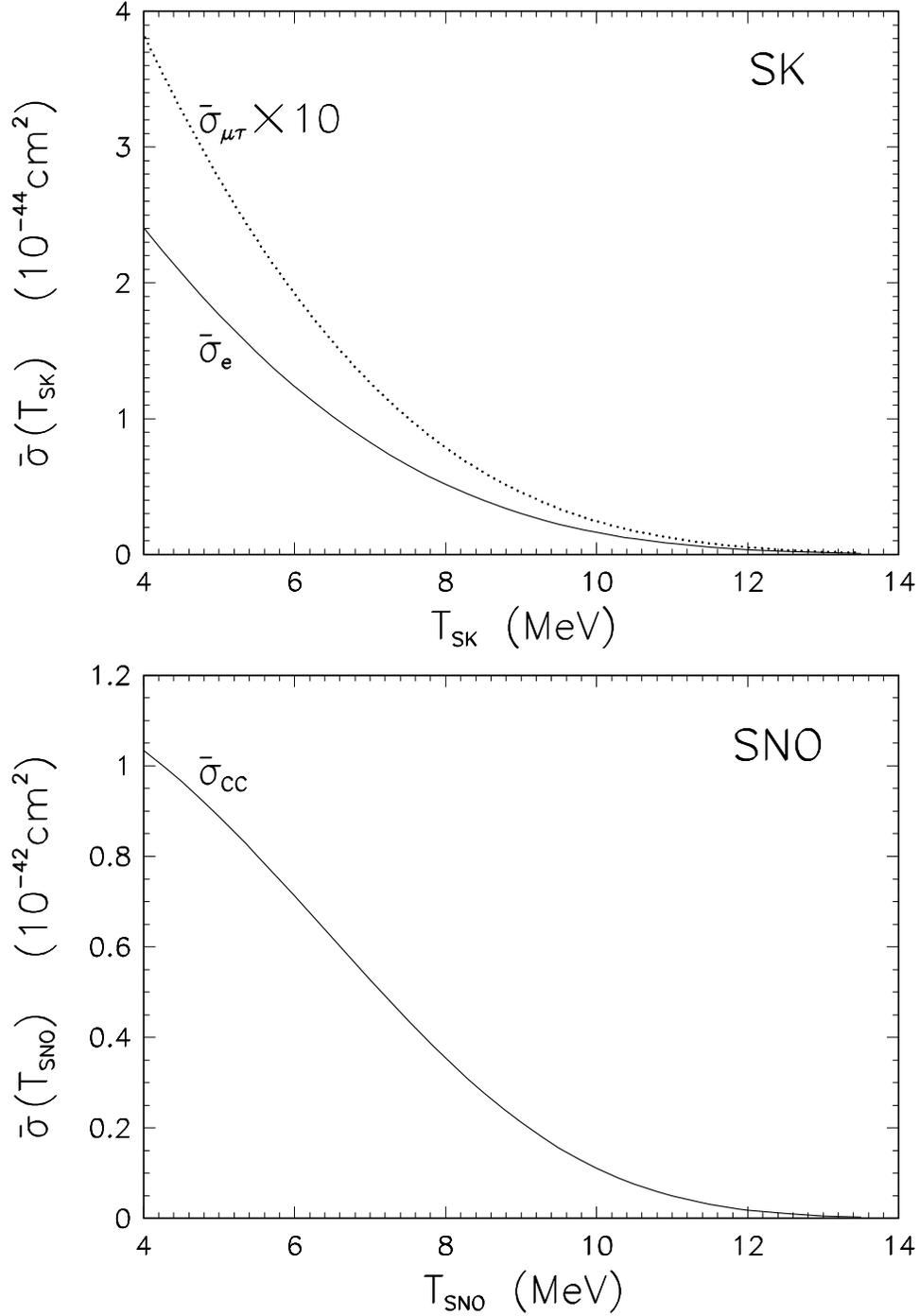

\caption{Upper panel: The  effective cross sections for 
$\nu_e\, e\to \nu_e\, e$ and
$\nu_{\mu,\tau}\,e\to \nu_{\mu,\tau}\,e$  scattering,
$\overline\sigma_e$ and $\overline\sigma_{\mu\tau}$, as a function
of the electron kinetic energy threshold in the SK experiment. 
Lower panel: The  effective cross section $\sigma_{\rm CC}$ for the 
charged current process 
$\nu_e\, d\to p\, p\, e$, as a function of the electron kinetic
energy threshold in the SNO experiment.}
\label{fig1}
\end{figure}

\begin{figure}
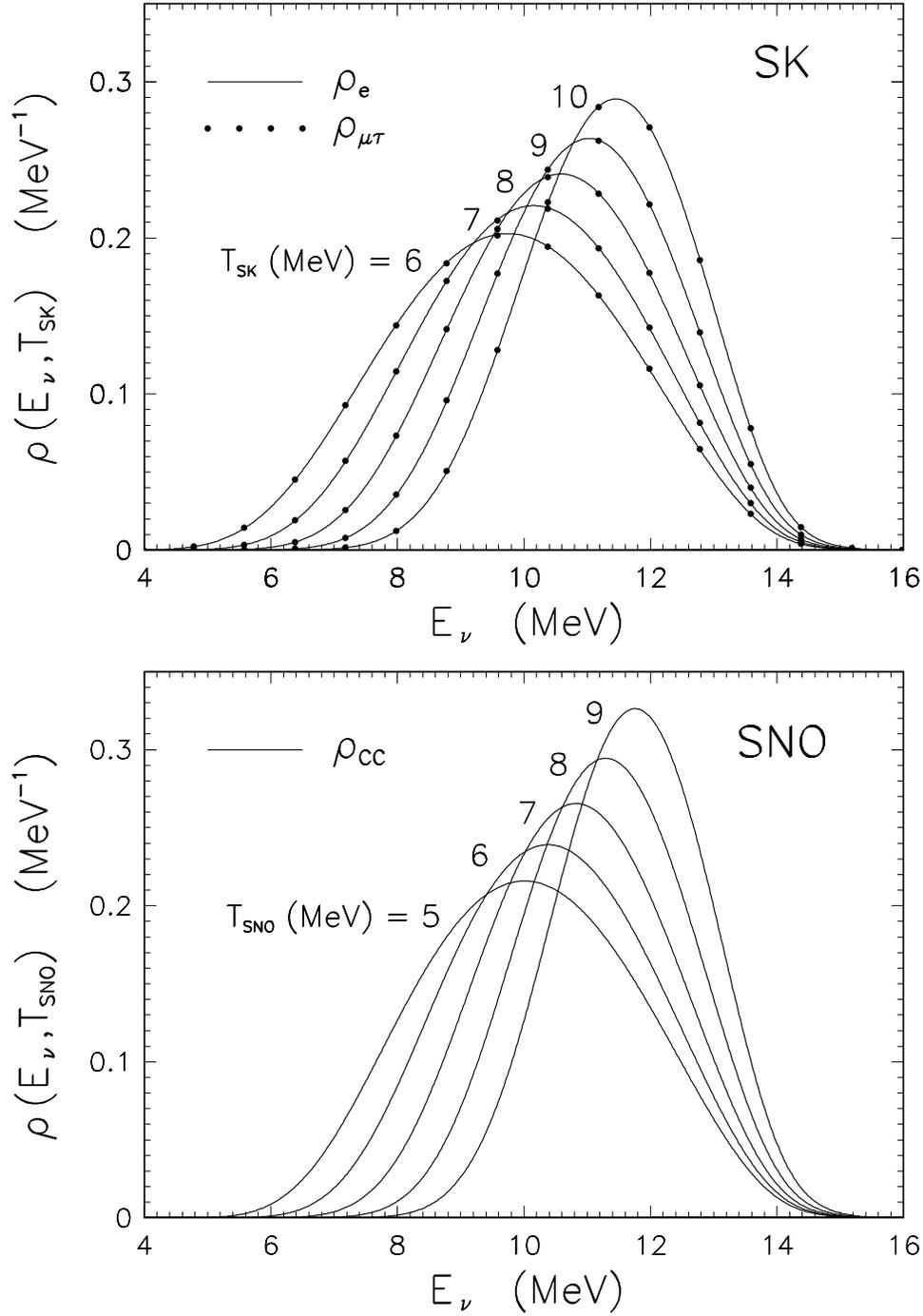

\caption{The normalized response functions of SK and SNO to
$^8$B neutrinos, for representative values of the detector
thresholds. See the text
for details.}
\label{fig2}
\end{figure}

\begin{figure}
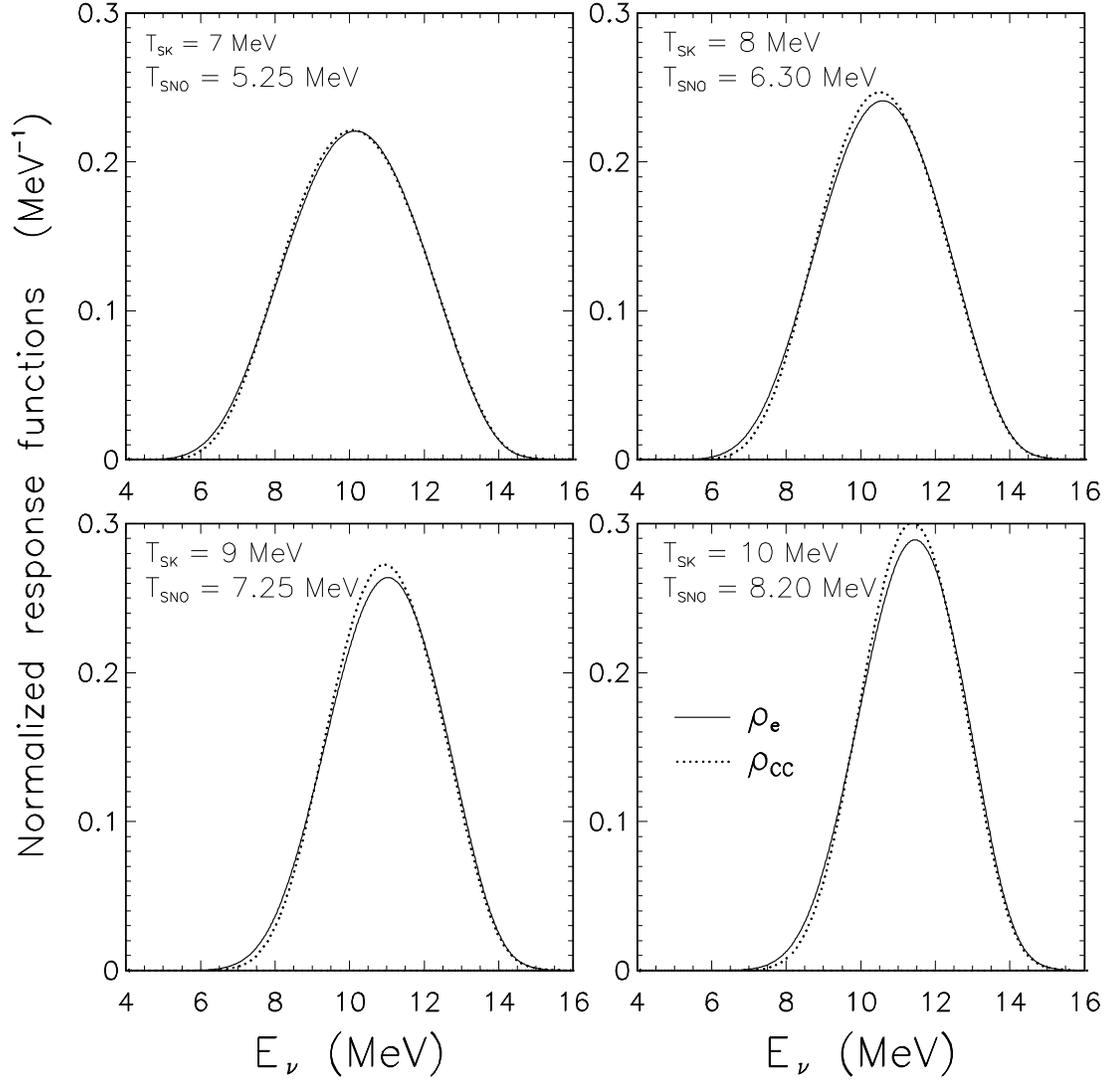

\caption{Examples of the approximate equality of the SK and SNO
response functions for selected values of the detector thresholds
$T_{\rm SK}$ and $T_{\rm SNO}$. Such values obey approximately the
relation $T_{\rm SNO} = 0.995\, T_{\rm SK} - 1.71$ (MeV).}
\label{fig3}
\end{figure}

\newcommand{\InsertFigure}[2]{\newpage\begin{center}\mbox{%
\epsfig{bbllx=1.4truecm,bblly=1.3truecm,bburx=19.5truecm,bbury=26.5truecm,%
height=21.4truecm,figure=#1}}\end{center}\vspace*{-1.8truecm}%
\parbox[t]{\hsize}{\small\baselineskip=0.5truecm\hspace*{0.5truecm} #2}}
\InsertFigure{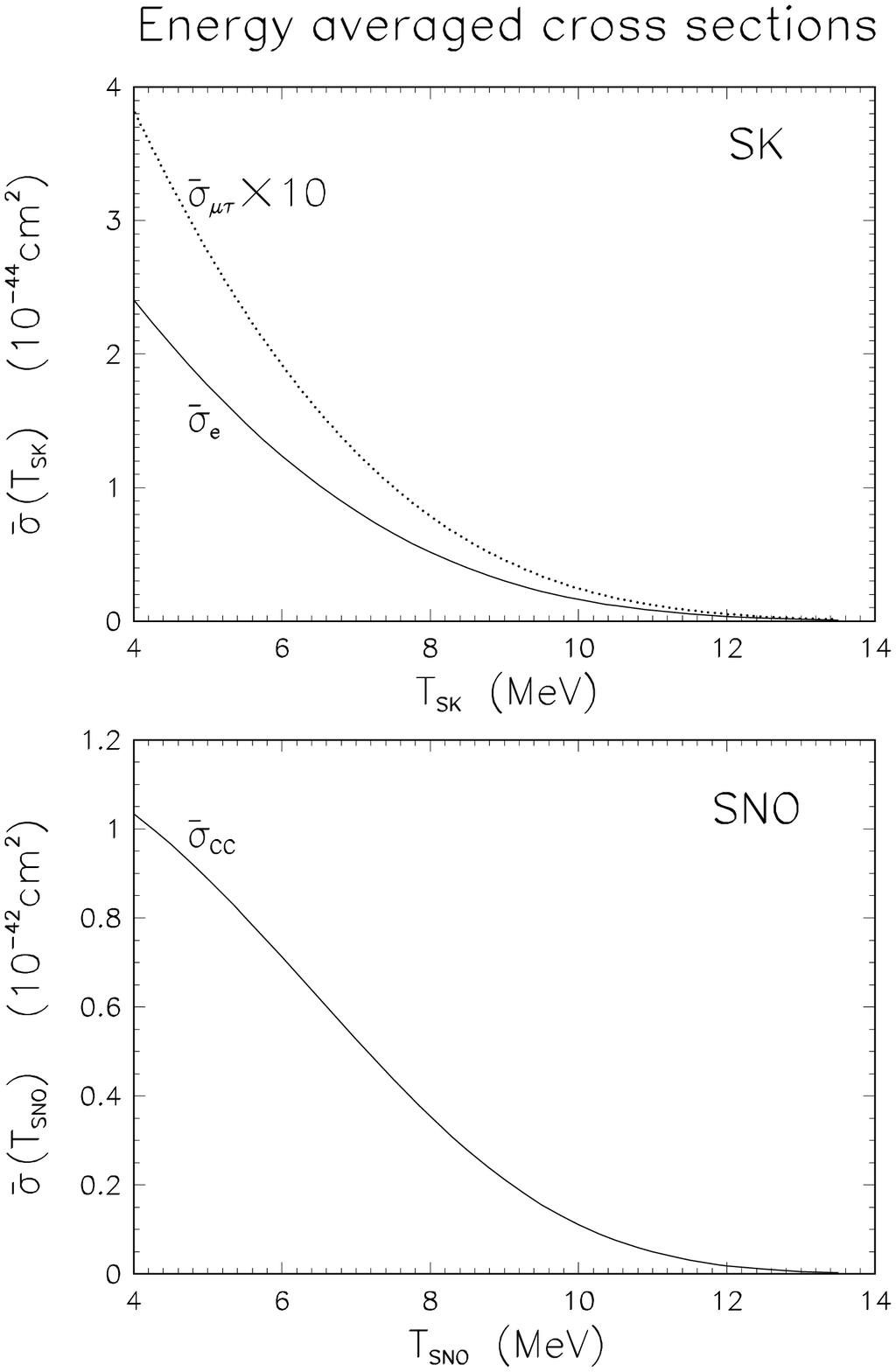}%
{   FIG.~1. Upper panel: The  effective cross sections for 
$\nu_e\, e\to \nu_e\, e$ and
$\nu_{\mu,\tau}\,e\to \nu_{\mu,\tau}\,e$  scattering,
$\overline\sigma_e$ and $\overline\sigma_{\mu\tau}$, as a function
of the electron kinetic energy threshold in the SK experiment. 
Lower panel: The  effective cross section $\sigma_{\rm CC}$ for the 
charged current process 
$\nu_e\, d\to p\, p\, e$, as a function of the electron kinetic
energy threshold in the SNO experiment.}
\InsertFigure{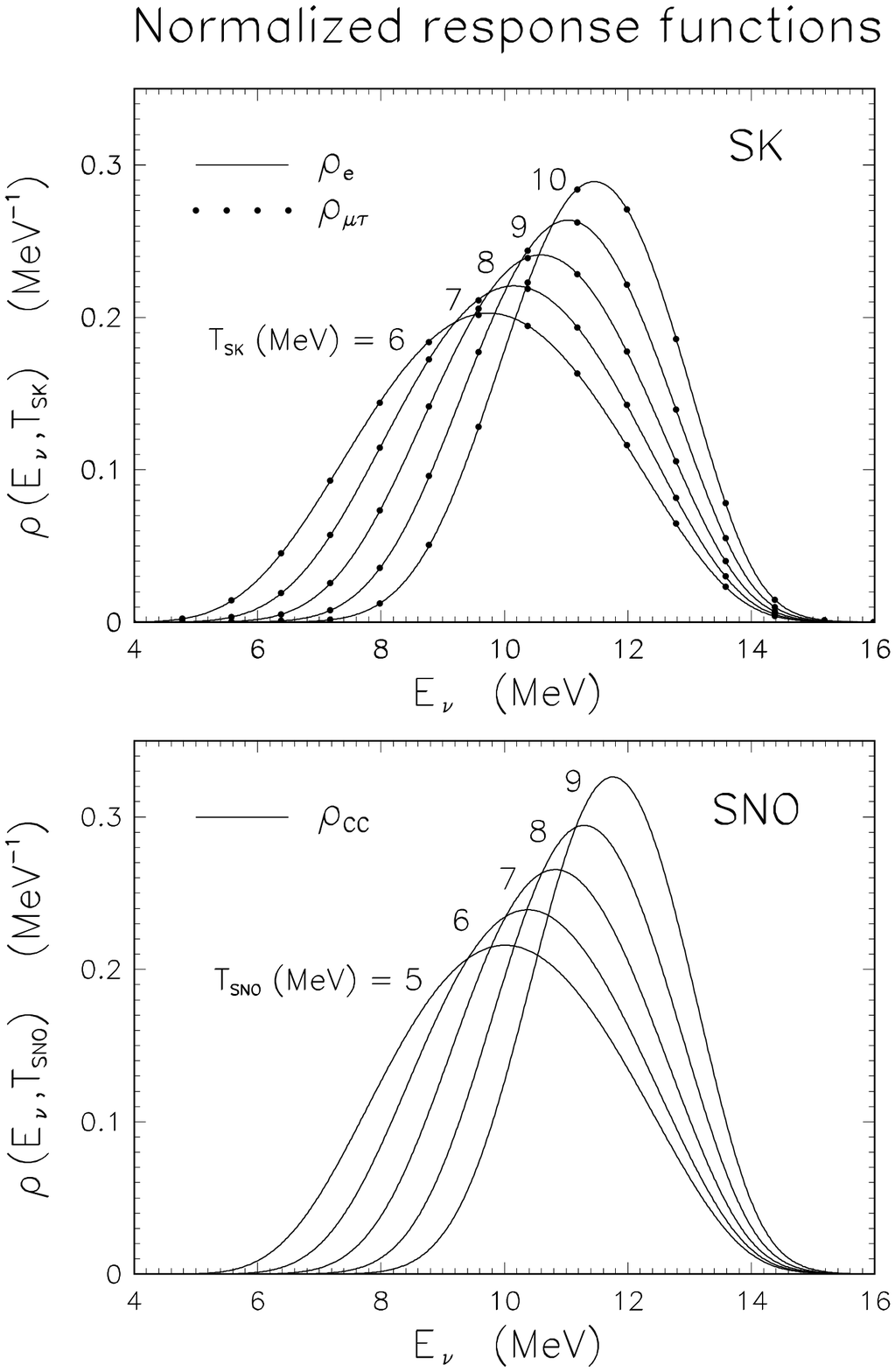}%
{   FIG.~2. The normalized response functions of SK and SNO to
$^8$B neutrinos, for representative values of the detector
thresholds. See the text
for details.}
\InsertFigure{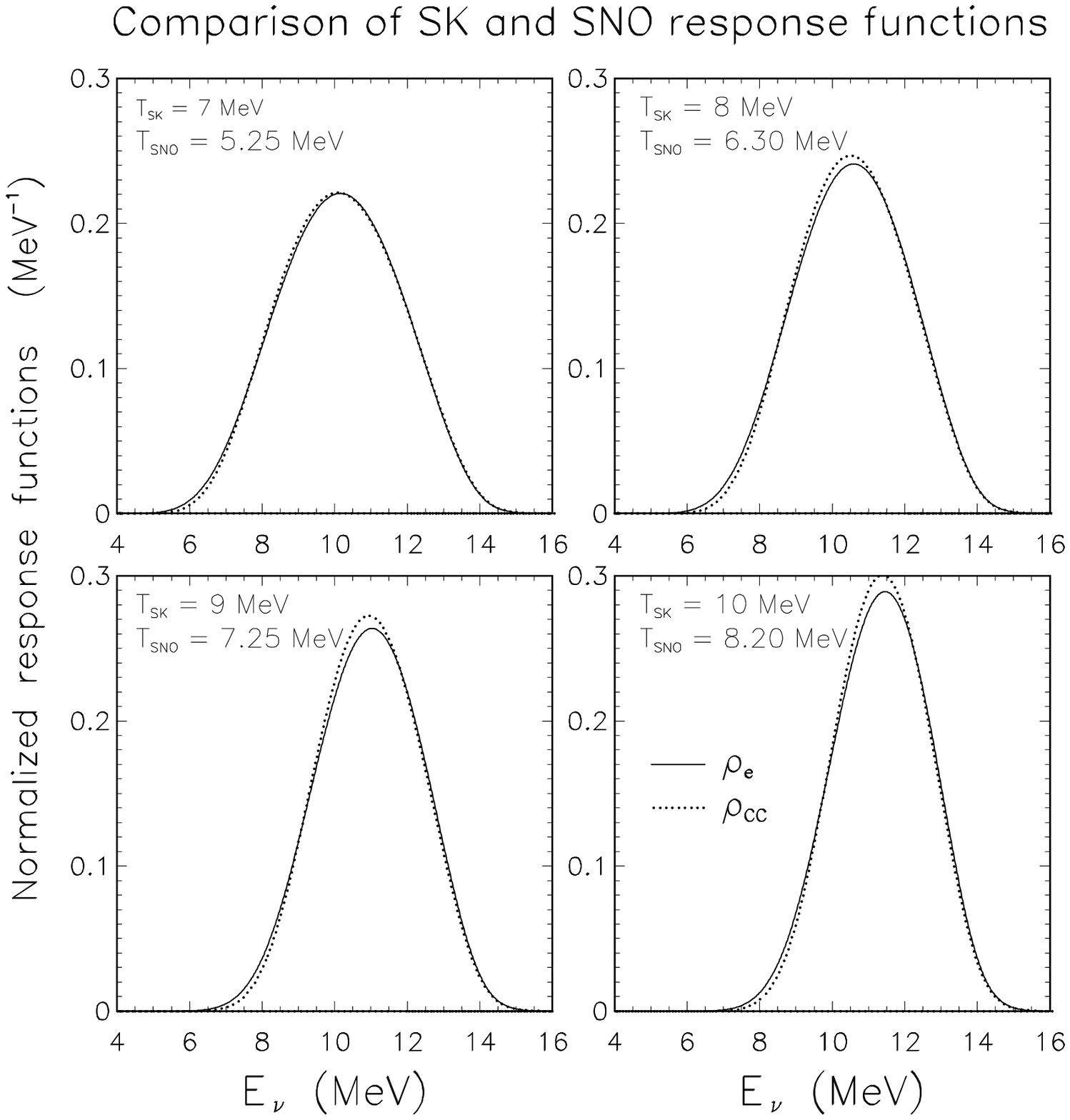}%
{   FIG.~3. Examples of the approximate equality of the SK and SNO
response functions for selected values of the detector thresholds
$T_{\rm SK}$ and $T_{\rm SNO}$. Such values obey approximately the
relation $T_{\rm SNO} = 0.995\, T_{\rm SK} - 1.71$ (MeV).}

\eject

\begin{references}

\bibitem{Fu98}	Super-Kamiokande Collaboration, Y.\ Fukuda {\em et al.\/},
		Report hep-ex/9805201, submitted to Phys.\ Rev.\ Lett.
		
\bibitem{Ca97}	V.\ Castellani, S.\ Degl'Innocenti, G.\ Fiorentini,
		M.\ Lissia, and B.\ Ricci,
		Phys.\ Rep.\ {\bf 281}, 309 (1997).
		
\bibitem{Su98}	Y.\ Suzuki, in {\em Neutrino~'98}, 
		XVIII International Conference on Neutrino Physics
		and Astrophysics, Takayama, Japan, June 1998, to
		appear in the Proceedings. Scanned transparencies
		available at the URL 
		http://www-sk.icrr.u-tokyo.ac.jp/nu98/scan/index.html~.
		
\bibitem{Na98}	M.\ Nakahata, in {\em New Era in Neutrino Physics},
		Satellite Symposium after {\em Neutrino~'98}
		\protect\cite{Su98}, Tokyo Metropolitan University,
		Japan, June 1998, to appear in the Proceedings. Scanned 
		transparencies available at the URL 
		http://musashi.phys.metro-u.ac.jp/era\_index.htm~.
		
\bibitem{Ba89}	J.\ N.\ Bahcall, {\em Neutrino Astrophysics}
		(Cambridge University Press, Cambridge, England, 1989)
		
\bibitem{Mc98}	A.\ McDonald for the SNO Collaboration, in {\em Neutrino~'98}
		\protect\cite{Su98}. 

\bibitem{Ya98}	S.\ Yamaguchi, PhD thesis, University of Osaka, Jan.\ 1998,
		pp.~79 and 81; available at the URL
		http://www-sk.icrr.u-tokyo.ac.jp/doc/sk/pub~.
		
\bibitem{Ba96}	J.\ N.\ Bahcall, E.\ Lisi, D.\ E.\ Alburger,
		L.\ De Braeckeleer, S.\ J.\ Freedman, and J.\ Napolitano,
		Phys.\ Rev.\ C {\bf 54}, 411 (1996).
		
\bibitem{BP98}	J.\ N.\ Bahcall, S.\ Basu, and M.\ H.\ Pinsonneault,
		Report astro-ph/9805135, submitted to Phys.\ Lett.\ B.
		
\bibitem{SNOt}	SNO detector technical specification, available at the URL
		http://snodaq.phy.queensu.ca/sno/sno.html~.
		
\bibitem{Si95}	J.\ N.\ Bahcall, M.\ Kamionkowski, and A.\ Sirlin,
		Phys.\ Rev.\ D {\bf 51}, 6146 (1995).
		
\bibitem{Ku94}	K.\ Kubodera and S.\ Nozawa, 
		Int.\ J.\ Mod.\ Phys.\ E {\bf 3}, 101 (1994).
		We use the computer programs described in 
		J.\ N.\ Bahcall and E.\ Lisi, 
		Phys.\ Rev.\ D {\bf 54}, 5417 (1996)
		and available at the URL
		http://www.sns.ias.edu/$^\sim$jnb/SNdata~.
		
\bibitem{Hi87}	S.\ Hiroi, H.\ Sakuma, T.\ Yanagida, and M.\ Yoshimura,
		Phys.\ Lett.\ B {\bf 198}, 403 (1987).
		
\bibitem{Ba91}	V.\ Barger, R.\ J.\ N.\ Phillips, and K.\ Whisnant,
		Phys.\ Rev.\ D {\bf 43}, 1110 (1991).
		
\bibitem{Kw94}	W.\ Kwong and S.\ P.\ Rosen,
		Phys.\ Rev.\ Lett.\ {\bf 73}, 369 (1994).
		
\bibitem{Kw95}	W.\ Kwong and S.\ P.\ Rosen,
		Mod.\ Phys.\ Lett.\ A {\bf 10}, 1331 (1995),
		Phys.\ Rev.\ D {\bf 54}, 2043 (1996).
%
%
\bibitem{Ch85}	H.\ H.\ Chen,
		Phys.\ Rev.\ Lett.\ {\bf 55}, 1534 (1985).


\end{references}
\end{document}